\documentstyle[prl,aps,twocolumn,epsf]{revtex}

\newcommand{\be}{\begin{equation}}
\newcommand{\ee}{\end{equation}}
\newcommand{\beqa}{\begin{eqnarray}}
\newcommand{\eeqa}{\end{eqnarray}}
\newcommand{\bean}{\begin{eqnarray*}}
\newcommand{\eean}{\end{eqnarray*}}
\newcommand{\eqn}[1]{(\ref{#1})}

\newcommand{\del}{\partial}
\newcommand{\nn}{\nonumber}
\def\tr{\mathop{\mbox{Tr}}\nolimits}
\newcommand{\snabla}{\nabla\!\!\!\!/}
\def\up#1{\leavevmode \raise.16ex\hbox{#1}}

\def\thebibliography#1{{\bf REFERENCES\markboth
 {REFERENCES}{REFERENCES}}\list
 {[\arabic{enumi}]}{\settowidth\labelwidth{[#1]}\leftmargin\labelwidth
 \advance\leftmargin\labelsep
 \usecounter{enumi}}
 \def\newblock{\hskip .11em plus .33em minus -.07em}
 \sloppy
 \sfcode`\.=1000\relax}

\begin{document}
\title{Critical exponents of the Gross-Neveu model from the effective 
average action}
\author{L.Rosa, P. Vitale 
\thanks{Alexander von Humboldt fellow} 
and C. Wetterich }
\address{
$^1$ Institut f\"ur Theoretische Physik, Universit\"at
Heidelberg, Philosophenweg 16, D-69120, Heidelberg Germany}

\maketitle

\small{ HD-THEP-00-30}

\begin{abstract}
{The phase transition of the Gross-Neveu model with $N$
fermions is investigated by means of a non-perturbative evolution
equation for the scale dependence of the effective average action. The
critical exponents and scaling amplitudes are calculated for
various values of $N$ in $d=3$. It is also explicitely verified that
the Neveu-Yukawa model belongs to the same universality class as the
Gross-Neveu model.} 
\end{abstract}

\narrowtext


\bigskip

The Gross-Neveu (GN) model \cite{gn}
is one of the simplest models for interacting fermions.
Nevertheless, in three dimensions our quantitative understanding beyond 
some universal characteristics of the phase transition has remined rather 
incomplete. The universality class of the GN-model in
dimensions between 2 and 4 has been argued to be the same as 
the Neveu-Yukawa (NY) model \cite {ZJ} in $4-\epsilon$  dimensions 
 \cite{ZJ2}. Both the large $N$ and
$\epsilon$ expansion indicate that a second order phase transition
takes place for some critical value of the coupling constant if the 
number of 
fermion species $N$ is larger than one \cite{park}. The anomalous
dimensions have been   calculated up to the third order in the  $ 1/N
$  expansion \cite{gracey}, while some critical exponents have been
computed to the order $1/N$ in the phase with spontaneous symmetry 
breaking (SSB) \cite{kogut}. In this letter we find the second order phase
transition and calculate the critical exponents employing an analytical
method based on nonperturbative flow equations for scale dependent
effective couplings. We directly obtain results for arbitrary dimension
and without a restriction to large $N$.  Despite the presence of 
massless fermions we are able to investigate the symmetric phase. 
Due to the fermion fluctuations the infrared physics is not trivial in the 
NY-language and requires a careful discussion of the critical exponents. 
Beyond the universal critical behaviour our method gives a description for 
arbitrary values of the GN-coupling away from the critical point. In 
particular, we compute the non-universal critical amplitudes.

The running couplings parameterize the
effective average action $\Gamma_k$ \cite{wett1} which is a type of
coarse grained free energy. It includes the effects of the quantum
fluctuations with momenta larger than an infrared cutoff $~k$. In the
limit where the average scale $k$ tends to zero $\Gamma_{k}$ becomes
therefore the usual effective action, i.e. the generating functional of
$1PI$ Green functions. In the limit $k\rightarrow \infty$ it approaches
the classical action. In a theory with bosons and fermions the scale
dependence of $\Gamma_k$ can be described by an exact nonperturbative
evolution equation \cite{wett1,wett2}
\beqa
{\del\over\del t} \Gamma_k[\phi,\psi]&=& {1\over 2} \tr \left\{
(\Gamma_k^{(2)}[\phi,\psi] +{\cal R}_k)^{-1}_{B} {\del\over\del t}  
R_{kB}\right.\nn\\
&-& \left. (\Gamma_k^{(2)}[\phi,\psi] +{\cal R}_k)^{-1}_{F} {\del\over\del
t} 
R_{kF} \right\} \label{1.1}
\eeqa
where $t=\ln(k/\Lambda)$ with $\Lambda$ some suitable high momentum scale.
The trace represents a momentum integration as well as a summation
over internal indices and $\Gamma_k^{(2)}$ is the exact inverse propagator
given by the matrix of second functional derivatives of the action with
respect to bosonic and fermionic field variables. The infrared cutoff
$$
{\cal R}_k(q,q')=
\left(
\begin{array}{ccc}
R_{kB}&0&0\\
0&0&R_{kF}\\
 0&-R_{kF}&0
\end{array}
\right)
(2\pi)^d \delta^d(q-q')
$$
is parameterized by the bosonic and fermionic cutoff functions
${R}_{kB}(q)=q^2{Z_{\sigma,k} r_B(q)},~ 
{R}_{kF}(q)=i\not{q} Z_{\psi,k}r_F(q)  $. 
We choose
\be
R_{kB}=  \frac{Z_{\sigma,k} q^2}{e^{q^2\over k^2}-1 };~
R_{kF}= i Z_{\psi,k} \not{q}\left(\frac{1}{\sqrt{1-e^{-{q^2\over k^2}}}}
- 1 \right) 
\ee
where $Z_{\sigma,k},~Z_{\psi,k}$ are wave function renormalizations.
The momentum integration in Eq. \eqn{1.1} is both infrared and ultraviolet
finite. Equation \eqn{1.1} is an exact but complicated functional
differential equation wich can only be solved approximately by
truncating the most general form of $\Gamma_k$. Once a suitable
nonperturbative truncation is found the flow equation can be integrated
from some short distance scale $\Lambda$, where $\Gamma_\Lambda$ can be
taken as the classical action, to $k\rightarrow 0$ thus solving the
model approximately.

The GN model  is described in terms of a $O(N)$
symmetric action for a set of $N$ massless Dirac fermions. The
classical Euclidean action is given by 
\be 
S_{GN}=\int d^d x~
\left[{-\bar\psi}_{ i}(x) (\snabla + \sigma(x)) \psi_i(x) 
+{1\over 2 \bar G} \sigma^2(x)\right].
\ee
(Here and in the following we distinguish with a bar the dimensionful 
couplings.) The (pseudo)-scalar $\sigma(x)$ is an auxiliary non
dynamical
field which can be integrated out from the partition function,
leading to the replacement 
$\sigma(x)\rightarrow\bar{G}\bar{\psi}(x)\psi(x)$. 
Its vacuum expectation value $ \sigma_0$ is proportional to the fermion 
condensate, $ \sigma_0= \bar{G}<\bar \psi \psi>$.  The model is
asymptotically free and perturbatively renormalizable in 2 dimensions,
hence it exhibits a non-trivial  fixed point in $d=2+\epsilon$.
It is $1/N$ renormalizable in $2<d<4$. 

The NY model  whose classical action is
\beqa 
S_{NY}&=& \int d^d x~
\left[-{\bar\psi}_{ i}(x) (\snabla + \bar h \sigma(x)) \psi_i(x) 
+  {1\over 2}(\del_\mu \sigma(x))^2 \right.\nn\\
&+& \left.{m^2\over 2} \sigma^2(x) + {\bar g\over 4!} \sigma^4(x) \right]
\eeqa
has a Gaussian fixed point in $d=4$ where it is perturbatively 
renormalizable and a non-trivial fixed point in $d=4-\epsilon$. 
Both  models have, in even dimensions, a discrete chiral symmetry 
which prevents from the addition of a fermion mass term, while in odd 
dimensions a mass term is forbidden by  space parity.  
Performing a large $N$ analysis the universal properties of the two models 
are argued to be the same in $2<d<4$ \cite{ZJ2}:in such limit the two 
models are equivalent in the scaling region if we rescale $\bar h 
\sigma$ to $\sigma$ and set $\bar G={\bar h}^2/m^2$.  

We consider a truncation of the effective action $\Gamma_k$ which
contains a potential for the scalar field and a Yukawa term.
 In momentum space it reads ($\int dq=\int d^d q/(2\pi)^d$)
\beqa
&&\Gamma_k[\sigma,\psi,\bar\psi]=
\int d^d x U_k(\sigma) +  \int d q \left[{Z_{\sigma,k}\over 2} 
 \sigma(-q) q^2 \sigma(q) \right.\nn\\
&&\left. - Z_{\psi,k} \bar\psi_i(-q)i\not{q}\bar\psi^i(q)
 -\int d p~ 
\bar{h}_k\bar\psi_i(-q)\sigma(p)\psi^i(q-p)\right].
\label{gammaq}
\eeqa
The scalar potential is assumed to be a function of the invariant 
$\sigma^2(x)$ and we make the further simplification
\beqa
U_k(\sigma) &=& {{m}^2_k\over 2}(\sigma^2(x)-\sigma_{0k}^2) + 
{\bar{g}_k\over 4!} (\sigma^2(x)-\sigma_{0k}^2)^2  \nn \\
 &+& {\bar{b}_k\over 6!} (\sigma^2(x)-\sigma_{0k}^2)^3 .
\label{us}
\eeqa
The symmetric regime is characterized by the minimum being at 
$\sigma_{0k}^2=0$. In the SSB regime a $k$-dependent minimum 
$\sigma^2_{0k}\ne 0$ develops, whereas ${m}_k^2=0$.


Inserting Eqs. \eqn{gammaq}, \eqn{us} into \eqn{1.1} we 
obtain a set of evolution or renormalization group equations (RGE) for the
effective parameters of the theory in the two regimes.
 Integrating the RGE between some high 
momentum scale $\Lambda$ and $k=0$ we will find the phase
transition point and extract the critical 
exponents  of the theory. 
We find it convenient to work with dimensionless quantities
$
h^2_k=Z_\sigma^{-1} Z_\psi^{-2} k^{d-4} \bar{h}_k^2~,~
g_k=Z_\sigma^{-2} k^{d-4} \bar{g}_k~,~
b_k=Z_\sigma^{-3} k^{2d-6} \bar{b}_k~,~
e_k=Z_\sigma^{-1} k^{-2} {{m}^2_k}~,~
{\tilde\rho}=\frac{1}{2}Z_\sigma k^{2-d} \sigma^2~,~
{\kappa}_k=\frac{1}{2}Z_\sigma k^{2-d} \sigma^2_{0k},~
u_k=U_k k^{-d}
$
and we use 
$u'_k=\frac{\partial u_k}{\partial \tilde\rho}$ etc.

The evolution equation for the potential 
obtains from \eqn{1.1} by evaluating $\Gamma_k^{(2)}$ in the truncation 
\eqn{gammaq} for a constant background scalar field. We find
\beqa   
{\del_t U_k(\sigma) \over k^d} &=&  v_d \int_0^\infty dy 
{y^{d/2}} \left\{ { -\eta_\sigma r_B-2y \dot{r}_B\over 
u'_k+2\tilde\rho u''_k + y(1+r_B) } \right.\nn\\
&+& 2 N' \left. { (\eta_\psi r_F+2y \dot{r}_F)(1+r_F)\over 
2 h^2_k\tilde\rho+ y(1+r_F)^2 } \right\}\equiv 
{\cal \zeta}_k(\tilde\rho)~. \label{evopot}
\eeqa
Here we have introduced the notation $N'= 2^{\gamma/2} N$ with
$2^{\gamma/2}$ the
dimension of the
$\gamma$ matrices and $y=q^2/k^2,~
\dot r=\frac{\partial r}{\partial y}$. Also we have defined $v_d^{-1}=
2^{d+1}\pi^{d/2}\Gamma(d/2)$.
In the SSB regime the evolution equations for the parameters 
$\kappa_k,~ g_k$ and $b_k$ are then obtained as
 \beqa
\del_t \kappa_{k} &=&(2-d-\eta_\sigma) \kappa_{k} - 
{3\over g_k}\left[\del_{\tilde\rho} {\cal \zeta}_k \right]_{\kappa_{k}}
 \label{evochi}\\
\del_t g_k&=&(2\eta_\sigma+d-4) g_k 
+3\left[\del^2_{\tilde\rho} {\cal \zeta}_k \right]_{\kappa_{k}}
+\frac{1}{5}b_k \del_t {\kappa_{k}} \\
\del_t b_k&=&(3\eta_\sigma+2d-6) b_k + 15\left[\del^3_{\tilde\rho}
 {\cal \zeta}_k \right]_{\kappa_{k}}~.
\eeqa
In the symmetric regime ${\kappa_{k}}=0$ so we replace
 eq. \eqn{evochi} by
\be
\del_t e_k=(\eta_\sigma-2) e_k + \left[\del_{\tilde\rho}{\cal \zeta}_k
\right]_0 ~.
\ee
The anomalous dimensions
$\eta_\sigma$ and $\eta_\psi$ are defined as 
 \be
\eta_\sigma (k) = - \del_t \ln Z_{\sigma,k}~,~~~
\eta_\psi (k) = - \del_t \ln Z_{\psi,k}~. 
\ee
The wave function renormalizations $Z$ parametrize the momentum dependence 
of the propagators at zero momentum and $\sigma=\sigma_{0k}$. One finds
\beqa
\eta_\sigma(k)&=& 
\del_\alpha \left\{  {v_d\over d}  \int dy~y^{d/2}
\left[( {2\over 15}b_k {\kappa_{k}}
+g_k)^2{\kappa_{k}}{\dot H(y,\alpha)}^2 \right.\right.\nn\\ 
&-& \left.\left.  2 N h^2_k \left( 2^\gamma y {\dot G(y,\alpha)}^2 
- 2{h^2_k} {\kappa_{k}}  {\dot 
F(y,\alpha)}^2\right) \right]\right\}_{\alpha=0} \label{etasig}\\
\eta_\psi(k)&=& 4  
h^2_k \del_\alpha \left\{{v_d\over d} \int dy~y^{d/2} {\dot H(y,\alpha)} 
G(y,\alpha)\right\}_{\alpha=0}~\label{etaps}
\eeqa
with
\beqa
H(y,\alpha)&=&{1\over e_k+ 
{2\over 3} g_k \kappa_k 
+y[(1+r_B)-\alpha(\eta_\sigma r_B +2 
y \dot r_B)]} \nn \\
G(y,\alpha)&=&{1+r_F-\alpha(\eta_\psi r_F + 2y \dot r_F)\over y[1+r_F
-\alpha(\eta_\psi r_F + 2y \dot r_F)]^2+2 h^2_k {\kappa_k} } \\
 F(y,\alpha)&=&{1\over y[1+r_F
-\alpha(\eta_\psi r_F + 2y \dot r_F)]^2+2 h^2_k {\kappa_k} } \nn
\eeqa
Finally, the evolution equation for the Yukawa coupling obtains
from 
taking derivatives of eq. \eqn{1.1} with respect to $\bar\psi,\psi$ and 
$\sigma$:
\beqa
\del_t h^2_k &=& (2\eta_\psi+\eta_\sigma+d-4)h^2_k\nn\\
&-& 4   h^4_k~
v_d \int_0^\infty dy~ y^{d/2} 
\Biggl[(\eta_\sigma r_B + 
2y \dot r_B) H^2(y,0) F(y,0)\nn\\
&-& 2  (\eta_\psi r_F + 
2y \dot r_F) {G^2(y,0) H(y,0)\over 1+r_F} \Biggr] \label{h2}.
\eeqa
The  equations \eqn{etasig},\eqn{etaps} and \eqn{h2} are valid 
in both regimes provided we set
$\kappa_k$ and $e_k$ appropriately.

If we expand this set of equations in the coupling constants we recover 
the one-loop results obtained in the $4-\epsilon$ expansion for the 
NY model \cite{ZJ}
\beqa
\partial_t g &=&-\epsilon g + {1\over 8\pi^2}\left({3\over 2} g^2 + 4Ngh^2-
24 N h^4\right)\\
\partial_t h^2 &=&-\epsilon h^2 + \frac{h^4}{8\pi^2}(2N+3)\\
\eta_\sigma&=&{Nh^2\over 4\pi^2},~~\eta_\psi={ h^2\over 16\pi^2}
\eeqa
Moreover, after identifying the running coupling constant of the GN model
as $G=h^2_k/e_k$ we also recover the one-loop result obtained in the 
$2+\epsilon$ expansion for the GN model \cite{ZJ}
\be
\partial_t G= (d-2)G - (N'-2){G^2\over 2\pi} +O(G^3). \label{G1loop}
\ee

We numerically evolve the flow equations \eqn{evochi}-\eqn{etaps} 
from a large momentum scale $\Lambda$ to $k\rightarrow 0$.
The initial values of the parameters are chosen in
such a way that $\Gamma_\Lambda=S_{GN}$:
$
Z_{\sigma\Lambda}\simeq 0,~Z_{\psi\Lambda}=1,~
\bar{h}^2_{\Lambda}=\Lambda,~\bar g_{\Lambda}=0,~\bar b_{\Lambda}=0.
$~
Then  $e_\Lambda=\left( Z_{\sigma\Lambda}G_\Lambda  \right)^{-1}$ 
is the only free parameter of the theory and plays the role of the
temperature. 
This value has to be tuned in
order to be near the second order phase transition. The value
corresponding  to the critical temperature $T_c$ is denoted by
$e_{\Lambda cr}$. 
For $Z_{\sigma \Lambda}=10^{-10}$, $d=3$ and $N=3$ we find in our 
truncation $e_{\Lambda cr}=1.878085212016 \cdot 10^9$ and the critical 
coupling in the GN model is $\bar{G}_{\Lambda cr}\Lambda=5.32$. 

The relevant parameter for the deviation form
$T_c$ is $\delta e= e_\Lambda- e_{\Lambda cr} = H(T-T_c)$ with constant
$H$.
 In Fig. 1 we show the behaviour of the dimensionless couplings as
functions of $k$ for $N=3$. As can be seen, they all reach a constant
value in the symmetric regime ($e_k>0$, $\kappa_k=0$) corresponding to a
nontrivial fixed point with vanishing beta functions. For $\delta e<0$
the symmetry is broken, as expected: the coupling $e_k$ goes to zero
and the order parameter $\sigma_{0k}$ acquires a non zero value. 
\begin{figure}[] \epsfysize=4cm
\epsfxsize=8cm
\epsffile{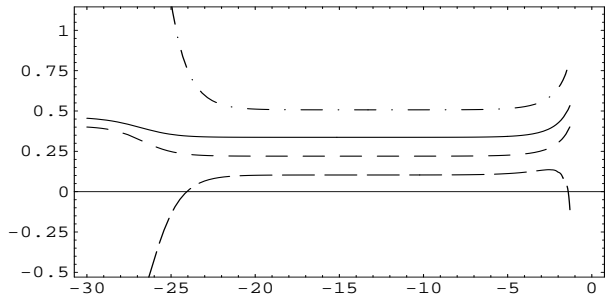}
\caption{ Running couplings $e_k, \frac{g_k}{100}, 
\frac{h^2_k}{10}$ and $\frac{b_k}{1000}$, as functions of 
$t=\ln k/\Lambda$ given by 
dot-dashed, full, dashed and long-dashed 
lines respectively.}
\end{figure} 
\noindent In the
following, whenever numerical results are reported,  we fix $d=3$ and
$N=3$. However, in Table 1. we summarize the results obtained for
$N=2,4,12$. 

In the symmetric (high $T$) phase the fermions
are massless. Their fluctuations induce a nontrivial dependence of
$Z_\sigma$ and the renormalized scalar mass $m_R^2(k)=Z^{-1}_\sigma(k)
m^2(k)$ on the scale $k$ even away from the phase transition. This
contrasts with the standard situation where the running of $m_R(k)$
essentially stops in the symmetric phase once $k$ becomes much smaller
than $m_R$. The issue of critical exponents in a situation with two
different infrared cutoffs $k$ and $m_R$ is therefore more complex than
usual. In a standard situation we would define the exponents $\gamma$
and $\nu$ by following the temperature dependence of the unrenormalized
and renormalized mass, $m^2(k)$ and $m_{R}^2(k)$
for $k\rightarrow 0$ \cite{wett3}.
Here we define the renormalized mass at some fixed small ratio 
$k/m_R$ by
\be
\bar{m}_R^2=m_R^2(k_c)-m_{R cr}^2(k_c),~~k_c= r_c \bar m_R \label{mren}
\ee
with $m^2_{R cr}(k)=e k^2$ on the critical trajectory. 
(In the numerical simulations we will fix the ratio $r_c$  to be
equal to 0.01).
This mass corresponds to the only relevant parameter characterizing
the critical behaviour. It is directly related to the deviation from the
critical temperature $T-T_c$ or $\delta e$. In the
following we use the arguments $\delta e$ or $\bar m_R$ interchangeably.
We also define the inverse susceptibility or unrenormalized mass by
\be
\bar{m}^2=\bar{m}_R^2 Z_\sigma(k_c, \bar{m}_R). \label{munren}
\ee 
Correspondingly, the critical exponents $\nu$ and $\gamma$ are defined for 
fixed $r_c$ and we find
\beqa
 \nu &=& {1\over 2} \lim_{\delta e\rightarrow 0} {\del
\ln\bar m_R^2(\delta e)\over \del
\ln\delta e}=1.041 \label{nu}\\
 \gamma &=& \lim_{\delta e\rightarrow 0}  
{\del \ln\bar{m}^2(\delta e)\over \del \ln\delta e}=1.323
\label{gamma}.
\eeqa
From the definition \eqn{munren} one has the relation 
\be
2\nu=\gamma -{\del \ln Z_\sigma(k_c,\bar m_R)\over \del\ln \delta e}.
\ee
A typical form of $Z_\sigma$ is 
\be
Z_\sigma\simeq Z_0\left({\bar m_R^2+k^2\over\Lambda^2}\right)^{-{1\over 
2}\bar\eta_\sigma}
 \left({k^2\over \bar m_R^2+k^2}\right)^{-{1\over 2}\eta_2}
\label{zetasigm}
\ee
and we conclude 
\be
{\del\ln Z_\sigma(k_c, {\bar m}_R)\over\del\ln\delta
e}=-\bar\eta_\sigma\nu,~~\gamma=\nu(2-\bar\eta_\sigma). \label{index3}
\ee
This is the usual index relation. We find 
$\bar\eta_\sigma= 0.729$ and the
 index relation \eqn{index3} yields $\gamma=1.323$, consistent with 
the value of $\gamma$ computed directly.
The index $\eta_2$, which vanishes in the standard situation,
determines the dependence of $\bar m_R$ on $r_c$.

For a more detailed understanding of the scale dependence we consider 
next the running of the renormalized mass and unrenormalized
mass with $k$. We define
\beqa
\hat\nu(k,\delta e) &=& {1\over 2} {{\del \ln[m_R^2(k,\delta e)-m_{R
cr}^2(k)]\over
\del
t}}|_{\delta e} \label{nuhat}\\
\hat\gamma(k,\delta e) &=&  {{\del \ln[m^2(k,\delta e)-\tilde{m}_{
cr}^2(k,\delta e)]\over \del
t}}|_{\delta e}  \label{gammahat}
 \eeqa
with 
$~
\tilde{m}^2_{cr}(k, \delta e)={m}^2_{R cr}(k) 
Z_\sigma(k, \delta e)~.
$
The relation $m^2_R=m^2/Z_\sigma$ implies the index relation
\be
\hat\gamma=2\hat \nu -\eta_\sigma \label{index1}
\ee
which differs from the usual relation $\gamma=\nu(2-\eta_\sigma)$.
For both $k$ and $\bar m_R$ sufficiently small and $k>> \bar m_R$ 
the indices $\hat \nu$, $\hat\gamma$,  $\eta_\sigma$ and $\eta_\psi$ 
approach constant values independent of
$k$ and $\bar m_R$ and we find 
\be
\hat\nu=0.502,~~\hat\gamma=0.295,~~\eta_\sigma=0.710,~~\eta_\psi=0.040.
\label{num1}
\ee
These values agree well with Eq. \eqn{index1}. In the opposite
regime, $k<<\bar m_R$, the running of the renormalized mass is only due to
the anomalous dimension $\eta_\sigma$ which is now different from the
value (\ref{num1}). We find 
$
\hat\nu=0.500,~~\hat\gamma=0.000,~~\eta_\sigma=1.000,~~\eta_\psi=0.000.
$
Again, these values agree well with Eq. \eqn{index1} and the 
expectation $\hat \gamma=0$. The nontrivial exponents $\hat\nu$, 
$\eta_\sigma$ in the NY-language correspond to the absence of 
renormalization effects for $\bar{G}_k$ for $k\rightarrow 0$ in the GN 
language.
 We note that for fixed $\delta e$ the renormalized scalar mass 
(which corresponds to the inverse 
correlation length) scales as $m_R\sim k$ 
for $m_R <<k$ and $m_R\sim \sqrt{k}$ for $m_R>>k$. 
The value of $\eta_\sigma=1$ for $k<<m_R$, 
corresponds to $\eta_2$ in eq. \eqn{zetasigm}.

We next turn to the dependence on the deviation $\delta e$ 
at fixed $k$. 
We define the exponents as
\beqa
\tilde \nu(k,\delta e) &=& {1\over 2} {{\del \ln[m_R^2(k,\delta e)-m_{R
cr}^2(k)]\over \del
\ln\delta e}}|_{k} \label{nutilde}\\
\tilde \gamma(k,\delta e) &=&  {{\del \ln[m^2(k,\delta e)-\tilde{m}_{
cr}^2(k,\delta e)]\over \del \ln\delta e}}|_{k}.  \label{gammatilde}
\eeqa
In the limit $\delta e \rightarrow 0$ we find 
$\tilde \nu = 0.520,~~\tilde\gamma=1.326$.
 The indices \eqn{nuhat},\eqn{gammahat} and
\eqn{nutilde},\eqn{gammatilde} are related to the  definition of the 
critical exponents \eqn{nu}, \eqn{gamma} by
\beqa
 \nu &=& \lim_{\delta e\rightarrow 0} (\tilde\nu (k_c,\delta 
e) +\hat\nu(k_c,\delta e)\nu)
\label{nu2}\\
 \gamma &=& 
\lim_{\delta e\rightarrow 0} (\tilde\gamma (k_c,\delta 
e) +\hat\gamma(k_c,\delta e)\nu)
\label{gamma2}
\eeqa
Here $k_c(\delta e)$ is given by evaluating \eqn{mren} at fixed $r_c$.
Equations \eqn{nu2} and \eqn{gamma2} yield respectively 
$\nu=1.041$ and $\gamma=1.326$, consistent with the results of  
\eqn{nu} and \eqn{gamma}. 

We also have computed the (nonuniversal) critical amplitudes which
describe the dependence of $\bar m_R$ and $\bar m$ on the coupling
$G_\Lambda$ of the GN model.  
Observing $\delta e/e_{\Lambda cr}=G_{\Lambda
cr}\delta(1/G_\Lambda)$ we obtain, for  small deviations from criticality,
\be
\bar{m}_R=A_\nu
\left|{ \delta G_{\Lambda }\over G_{\Lambda cr} }\right|^{\nu},~~ 
\bar{m}^2=A_\gamma \left|{ \delta G_{\Lambda }\over G_{\Lambda cr} }
\right|^{\gamma}.
\ee
Our numerical values of the amplitudes $A_\nu,~A_\gamma$ are 
 $A_\nu/\Lambda=0.016,~A_\gamma/\Lambda^2=0.212$.

We finally have investigated the low temperature phase with spontaneous  
symmetry breaking. Here the running of $\sigma_0$ stops for small $k$  
and the complications of the symmetric phase are absent. 
The critical exponent $\beta$ is defined with 
$\sigma_0=\lim_{k\rightarrow 0} \sigma_{0k}$ 
\be
\beta={1\over 2}\lim_{\delta e\rightarrow 0} {\ln\sigma_0^2 \over 
\ln\delta e}
\ee
such that
\be
\sigma_0= A_{\beta}
\left| {\delta G_{\Lambda }\over G_{\Lambda cr} }\right|^\beta~.
\ee
We find $A_\beta/\sqrt{\Lambda}=0.008$, $\beta=0.903$, in good agreement 
with the 
scaling  relation
$\beta={\nu\over2}(d-2+\eta_\sigma)$. 
In Fig. 2 we plot the condensate $\sigma_0$ as a function of 
$G_{\Lambda }$.
\begin{figure}[] \epsfysize=4cm
\epsfxsize=8cm
\epsffile{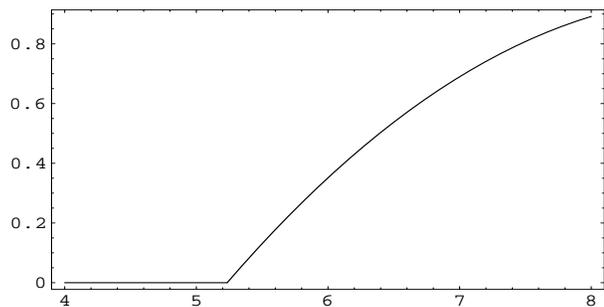}
\caption{Fermion-antifermion condensate $\sigma_0/\sqrt\Lambda$ as a 
function of the $(\bar\psi\psi)^2$ coupling $G_\Lambda$, in the range 
$[3/4 G_{\Lambda cr},3/2 G_{\Lambda cr}]$.}  
\end{figure} 
In  the Table we report our results for different values of $N$. For
all $N\geq2$ the 
existence of a second order phase transition is confirmed  
by our analysis. As can be checked, the scaling relations are  
well verified. To compare with existing results obtained in the $1/N$ 
expansion, let us fix $N =12$. In \cite{kogut} the critical exponents 
have been calculated to the order $1/N$:
\beqa
\nu &=& 1+{8\over 3N\pi^2 }=1.022 ,~~~\gamma=1+{8\over N\pi^2 }=1.068,\nn\\
\beta &=& 1+O({1\over N^2 }), ~~~~~~~~~~
\eta_\sigma=1-{16\over 3N\pi^2 }=0.955.
\eeqa    
In the same paper Montecarlo simulations for $N\ge 12$ are also reported. 
Conformal techniques  have been  used to calculate  the anomalous  
dimensions to $O(1/N^3)$  \cite{gracey} and yield,  for $N=12$,
$\eta_\psi=0.013,\eta_\sigma=0.913$. 
However,such techniques rely on being exactly at the critical 
point  and hence cannot be used to calculate the other exponents.
\begin{center}
\begin{tabular}{|c||c|c|c|c|} \hline
N & 2 & 3 & 4 & 12 \\ \hline
$\nu$    &~ 0.961~ &~ 1.041~ &~ 1.010~ &~ 1.023~ \\ \hline
$\gamma$ & 1.384 & 1.323 & 1.228 & 1.075 \\ \hline
$\nu(2-\bar\eta_\sigma)$ & 1.403 & 1.323 & 1.230 & 1.075 \\ \hline 
$\beta$  & 0.745 & 0.903 & 0.910 & 0.998 \\ \hline
$\frac{\nu}{2}(1+\eta_\sigma)$ & 0.750  
& 0.890  & 0.903  & 0.991 \\ \hline
$A_\nu/\Lambda$    & 0.007 & 0.016 & 0.009 & 0.014 \\ \hline
$A_\gamma/\Lambda^2$ & 0.042 & 0.212 & 0.233  & 0.968 \\  \hline
$A_\beta/\sqrt{\Lambda}$  & 0.007 & 0.008 & 0.005 & 0.007 \\  \hline
$\eta_\sigma$ & 0.561 & 0.710 & 0.789 & 0.936 \\  \hline
$\eta_\psi$   & 0.066 & 0.040 & 0.027 & 0.007 \\  \hline
$\bar\eta_\sigma$ & 0.541 & 0.729 & 0.765 & 0.971 \\  \hline
$G_{\Lambda cr}$ &9.989 &5.325 &3.613 &1.006 \\ \hline
\end{tabular}

\medskip

{TABLE 1. Critical exponents and amplitudes 
for different values of N~~~~~~~~~~~~~~~~~~~~~~~~~~~~
~~~~~~~~~~~~~~~~~~~~~~~~~~~~~~~~~~~~~~~}
\end{center}
The case $N=1$ appears to be different from $N>1$. We find a phase 
transition. For small $G_\Lambda$ (large $e_\Lambda$) eq. \eqn{G1loop} is 
valid 
($N'=2$) and $G_k$ scales according to its canonical dimension. The model 
is in the symmetric phase. For $G_\Lambda>G_{\Lambda cr}$, 
$G_{\Lambda cr}=19.416$ the mass term 
at the origin of the potential becomes negative, indicating spontaneous 
symmetry breaking. We find no scaling solution, neither for $e_k\ge 0$ 
nor for $\kappa_k\ge 0$. This may suggest a first order 
transition.

In conclusion, a simple truncation of the exact flow equation for the  
effective average action gives a consistent picture for a second order 
phase transition for the GN model with $N\ge 2$ in three dimensions. We 
have computed critical exponents and amplitudes and we relate directly 
physical observables like the correlation length or the order parameter to 
the value of the coupling $G$. By choosing different initial conditions we 
have also explicitely verified that the Neveu-Yukawa model belongs to the 
same universality class as the Gross-Neveu model. 
The universal exponents are
independent of the initial parameters, but not quantities like
$\sigma_0$, the renormalized mass and the corresponding amplitudes.

\end{document}